\documentclass{PoS}

\title{More effective theory of nuclear forces}

\ShortTitle{More effective theory of nuclear forces}

\author{\speaker{Michael C. Birse}\\
School of Physics and Astronomy, The University of Manchester,
Manchester, M13 9PL, U.K.\\
E-mail: \email{mike.birse@manchester.ac.uk}}

\abstract{I outline why the renormalisation group is needed to analyse the 
scale dependence and hence determine the power counting for effective theories 
of strongly interacting systems. I summarise the results of several such 
analyses for two- and three-body forces in nuclear physics. These show that a 
number of terms should be significantly promoted relative to naive dimensional
analysis.}

\FullConference{6th International Workshop on Chiral Dynamics\\
		 July 6-10 2009\\
		 Bern, Switzerland}

\begin{document}

\section{The problem with building an EFT for nuclear forces}

Chiral perturbation theory \cite{wein79} is an effective field theory (EFT) 
that can provide a systematic expansion of hadronic observables in powers of 
ratios of low-energy scales $Q$ (momenta, $m_\pi$, \dots) to scales of the
underlying physics $\Lambda_0$ ($m_\rho$, $M_N$, $4\pi F_\pi$, \dots). Its 
terms are organised by naive dimensional analysis, or ``Weinberg power 
counting'', which simply counts powers of the low-energy scales. This theory 
is perturbative, as its name implies, working for weakly interacting systems 
such as low-energy pions, photons and up to one nucleon. In contrast, 
nucleons interact strongly at low-energies, forming bound states (nuclei). 
To describe them, some interactions must be treated nonperturbatively.

\begin{figure}[h]
\begin{center}
\includegraphics[width=2cm]{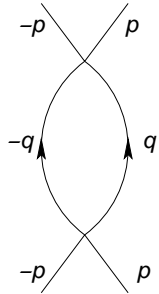}
\end{center}
\caption{Loop diagram for two-body scattering}
\end{figure}

To see why this might lead to a problem with extending the theory to two 
or more nucleons, look at the nonrelativistic loop diagram for NN scattering 
in Fig.~1. For contact interactions, this has the form
\begin{equation}
\frac{M}{(2\pi)^3}\int \frac{{\rm d}^3q}{p^2-q^2+{\rm i}\epsilon}
=-{\rm i}\,\frac{M\,p}{4\pi}+\mbox{analytic in}\ p^2.
\label{eq:basicloop}
\end{equation}
It is enhanced to order $Q$, instead of $Q^2$ as in the relativistic case. 
Nonetheless, the leading terms of the potential are of order $Q^0$ (OPE 
and the simplest contact interaction) \cite{wein90} and so each iteration 
is suppressed by a power of $Q/\Lambda_0$. The theory is therefore still
perturbative, provided $Q<\Lambda_0$.

In fact the analytic part of the integral (\ref{eq:basicloop}) is linearly 
divergent and so we need to either cut it off or subtract it at some scale 
$q=\Lambda$. Iterating the potential then leads to contributions with
powers of $\Lambda/\Lambda_0$. These will again be perturbative provided 
$\Lambda<\Lambda_0$, and so cannot generate bound states.

The commonly used workaround for this is the``Weinberg prescription'', where
people expand the potential to some order in $Q$ and then iterate it to all 
orders in their favourite dynamical equation (Schr\"odinger, 
Lippmann-Schwinger, \dots) \cite{wein90}. This has been widely applied (and 
even more widely invoked) in nuclear physics, but it leads to results for 
observables with no clear power counting. In particular, it resums a subset 
of terms to all orders in $Q$, some of which depend on the regulator but are 
of higher order than any of the terms kept in the potential. There are thus no 
counterterms to renormalise the results and cancel this regulator dependence. 
This need not be a problem provided these higher-order terms are small. 
But it is if we want to use these terms to generate bound 
states.

This issue has led to vigorous debate over the last twelve or more years and 
has left the nuclear EFT community polarised around two broad philosophies. 
One, the orthodox view, can be caricatured as ``The Prophet of EFT gave us 
the Power Counting in the holy texts, Phys Lett B251 and Nucl Phys B363.''
The other, liberal approach can be summarised as ``Let the renormalisation 
group decide!'' In terms of citations by the wider community, I must
admit that the orthodox party seems to be winning the election, at least 
so far. In this contribution, I shall try to summarise the liberal point 
of view.

\section{Renormalisation group}

The renormalisation group (RG) is a general tool for analysing scale 
dependence of systems in quantum or statistical mechanics. This means 
it can be used to determine the power counting that organises the 
effective theory describing a system. The RG was therefore an important 
piece of the motivation for the first chiral EFT, as noted by the Prophet 
himself \cite{wein09}.

The basic steps involved are as follows. First, we should identify all 
the relevant low-energy scales, $Q$. Of particular interest are any that 
promote leading-order terms to order $Q^{-1}$ since these can, and must, be 
iterated to all orders. (Iterations of these are not suppressed since 
each contributes a factor of order $Q^{-1}$ which cancels the $Q$ from 
the loop integration.) Examples of such scales in NN scattering are ones for 
the S-wave scattering lengths, $1/a\lesssim 40$~MeV \cite{vk98,ksw98}.
Then there is the ``unnatural'' strength of OPE, which is set by 
\begin{equation}
\lambda_{\scriptscriptstyle NN}
=\frac{16\pi F_\pi^2}{g_{\scriptscriptstyle A}^2 
M_{\scriptscriptstyle N}}\simeq 290\;\mbox{MeV}.
\end{equation}
This is built out of high-energy scales ($4\pi F_\pi$, 
$M_{\scriptscriptstyle N}$) but its magnitude is only $\sim 2m_\pi$, 
suggesting that it may be better treated as a low-energy scale.

Next, we cut off the theory at some arbitary scale $\Lambda$,
above the low-energy scales $Q$ but below the scale $\Lambda_0$
of the underlying physics, as in Fig.~2. (This assumes good separation 
of these scales, as required for a convergent expansion of the resulting 
EFT.)
\begin{figure}[h]
\begin{center}
\includegraphics[width=5cm]{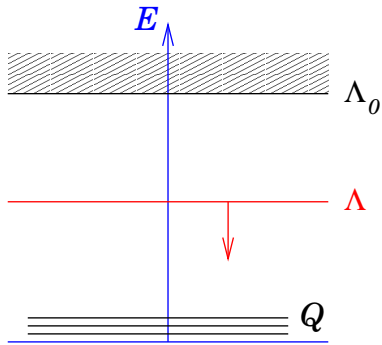}
\end{center}
\caption{The running cutoff $\Lambda$.}
\end{figure}

Then we can follow the evolution of our theory as we ``integrate out'' more 
and more of the physics by lowering $\Lambda$. (Don't even think about 
taking $\Lambda$ to infinity here!) As we vary this arbitrary cutoff,
we demand that physics (for example, the $T$ matrix) be independent of 
$\Lambda$.

Finally we rescale the theory by expressing all dimensioned quantities in 
units of $\Lambda$ so that we can look for fixed points of the RG flow---the 
$\Lambda$-independent end points that describe scale-free systems.
We can expand our EFT around one of these using perturbations that 
scale with definite powers of $\Lambda$. This scaling corresponds directly
to the order in the usual power counting, $\Lambda^\nu$ corresponding to a 
term in the EFT of order $Q^d$ where $d=\nu-1$. If you like, we have used 
$\Lambda$ to stand in for all our low-energy scales since it acts as the 
largest acceptable low-energy scale.\footnote{The RG identifies 
which terms in an EFT must be iterated to all orders, and which should be 
treated as perturbations. Despite my comments about $\Lambda$ above, we may 
use the resulting effective potential with a cutoff above $\Lambda_0$, but 
\emph{only} if we respect the counting the RG gives us. That is, we must iterate 
all terms of order $Q^{-1}$ and we must not iterate any ``irrelevant'' terms of 
order $Q^{+d}$. Otherwise we might, if we are very lucky, discover a new power 
counting, as in \cite{ntvk05}. But, more usually, we lose any consistent 
counting, as found by the many people who have tried iterating 
effective-range terms in short-range potentials, as in e.g.~\cite{pbc97}.}

\section{Fixed points of short-range forces}

Let me illustrate these ideas with the simplest example relevant to nuclear
physics: two particles interacting via short-range forces only (the pionless 
EFT). In this case we find two fixed points of the RG flow \cite{bmr98}.
One of them is the trivial fixed point $V_0=0$. Potentials near this point
describe weakly interacting systems. Their expansion around the fixed point
can be organised according to the usual perturbative (``Weinberg'') power 
counting.

Of more interest is the nontrivial fixed point. For a sharp momentum cutoff
this is
\begin{equation}
V_0(p,\Lambda)=-\,\frac{2\pi^2}{M\Lambda}\left[1-\frac{p}{2\Lambda}
\,\ln\frac{\Lambda+p}{\Lambda-p}\right]^{-1},
\end{equation}
where $p=\sqrt{ME}$ is the on-shell momentum. This potential is of 
order $Q^{-1}$ and so it must be iterated. It describes scattering in the
``unitary limit'', where the scattering length $a\rightarrow\infty$.
The expansion around this point has the form
\begin{equation}
V(p,\Lambda)=V_0(p,\Lambda)+V_0(p,\Lambda)^2\,\frac{M}{4\pi}\,\left(-\,\frac{1}{a}
+\frac{1}{2}\,r_e\,p^2+\cdots\right).
\end{equation}
The factor $V_0^2\propto \Lambda^{-2}$ multiplying each of the perturbations 
promotes them by two orders compared to naive expectations. This leads to the
``KSW'' power counting for systems with large scattering lengths 
\cite{vk98,ksw98}. The terms in the resulting expansion correspond directly to
the terms in the effective-range expansion.\footnote{This illustrates a general
feature of EFTs: the contact interactions are directly related to observables.
In this case the connection is to the phase shifts via the effective-range 
expansion. When long-range forces are included, the connection is either via
a DW Born expansion (for weakly interacting systems) or a DW effective-range
expansion (for strong short-range interactions) \cite{bb02}.}

Two-body scattering is sufficiently simple that we do not really need the full
power of the RG to understand the origin of this enhancement; it follows from
the form of the wave functions at short distances. Two particles in the unitary 
limit are described by  irregular solutions of the Schr\"odinger equation. 
At small radii (in $S$ waves) these behave as $\psi(r)\propto r^{-1}$.
Any cutoff smears a contact interaction over range $R\sim\Lambda^{-1}$.
We therefore need the extra factor of $\Lambda^{-2}$ in the interaction
in order to cancel the cutoff dependence from 
$|\psi(R)|^2\propto \Lambda^2$ in its matrix elements.

The same idea can also be used to understand the even stronger promotion
of three-body forces in systems of three bosons or three distinct fermions 
near the unitary limit (for example the triton). Here, naive dimensional 
analysis would suggest that the leading contact term is of order $Q^3$ 
\cite{wein90}. However, as the hyperradius $R\rightarrow 0$, the three-body 
wave functions have the form $\psi(R)\propto R^{-2\pm {\rm i}s_0}$, 
with $s_0\simeq 1.006$. The $R^{-2}$ divergence requires that the leading 
three-body force be promoted by four orders, to order $Q^{-1}$. The 
oscillatory behaviour associated with the imaginary part of the exponent 
is the origin of the Efimov effect \cite{ef71}. It means that the RG flow 
actually tends to a limit cycle instead of a fixed point \cite{bhvk98,bb04}.

\section{Effects of iterated one-pion exchange forces}

The central piece of OPE is the only one that contributes to scattering 
in spin-singlet waves. This has a $1/r$ singularity, which is not enough 
to alter the power-law forms of the wave functions at small $r$. The 
scattering in singlet waves with $L\geq 1$ is weak, and so the 
corresponding effective potential can be expanded using Weinberg power 
counting. In contrast, the $^1S_0$ channel has a low-energy virtual state. 
The expansion of its potential is like the one around the unitary fixed 
point, and can be organised using a KSW-like power counting.

The tensor piece of OPE is important in spin-triplet waves. It has a 
strong, $1/r^3$, singularity at the origin. The resulting short-distance
wave functions have the form $\psi(r)\propto r^{-1/4}$, multiplied by 
either a sine or an exponential function of 
$1/\sqrt{\lambda_{\scriptscriptstyle NN}r}$. As a result, short-range 
interactions are strongly promoted in these waves and so a new power 
counting needed, as observed by Nogga \textit{et al.}~\cite{ntvk05}. 
This gives the leading contact interaction an order $Q^{-1/2}$ 
in waves with $L=1$, 2 \cite{bi05,bi07}. The 
perturbative expansion for a potential of this order will converge very 
slowly and so it may be simpler to iterate it along with OPE.

Note that the importance of tensor OPE does depend on energy, and hence
of the cutoff $\Lambda$. At least at low energies, the centrifugal 
barrier protects the higher partial waves from probing the singular core
of the potential. Only waves above some critical momentum are able to
resolve singularity and hence require a nonperturbative treatment of OPE.
For waves with $L\geq 3$ this momentum is $p_c\gtrsim 2$~GeV and so
Weinberg power counting can be used for the usual choices of cutoff, 
$\Lambda\lesssim 600$~MeV. In contrast, waves with $L\leq 2$ have critical 
momenta $p_c\lesssim 3m_\pi$ and so NTvK counting is needed.\footnote{In 
the $^3S_1$ case, there is a further enhancement, similar to that in the
$^1S_0$ wave, associated with the deuteron bound state.}

Turning now to three-body forces, the two-pion exchange interaction is  
purely long-range and so it is not renormalised. Its leading 
contribution is thus of order $Q^3$, as in Weinberg's counting. Other 
contributions, involving two- or three-nucleon contact operators, are 
affected by the short-distance behaviour of the wave functions.

One-pion exchange interactions are represented by the diagram in Fig.~4.
The coefficients of the two-nucleon-one-pion contact operators are
generically denoted by $c_D$ \cite{engkmw02}.
\begin{figure}[h]
\begin{center}
\includegraphics[width=3cm]{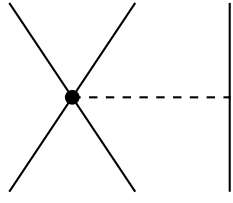}
\end{center}
\caption{A three-body OPE interaction.}
\end{figure}
The most important of these operators is the well-known one that couples 
the $^1S_0$ and $^3S_1$ NN channels \cite{hvkm00}. This is promoted to
order $Q^{5/4}$ by the nonperturbative treatment of the leading two-body
forces.\footnote{The wave function for a low-energy bound state 
in the $^3S_1$ channel is slightly less singular near the origin
than that in the $^1S_0$ channel, as a result of tensor OPE \cite{bi05}.}
In addition, there will be strong promotion of operators that couple $S$
and $P$ waves, and of those that couple various combinations of $P$ and
$D$ waves. 

Finally, we have the three-body contact interaction, whose strength is
denoted by $c_E$ \cite{engkmw02}. The order of this term
in the presence of tensor OPE is not known. I would expect it to be promoted,
but by less than in pionless case. Determining the counting for this
force will entail solving the three-body problem with $1/r^3$ potentials. 
This is in progress \cite{pl09}.

\section{A new road map for nuclear EFT}

\begin{table}[h]
\begin{center}
\begin{tabular}{l|c|c}
\hline
\ \ Order\ \ \quad & NN & NNN\\
\hline
\ \ $Q^{-1}$ & $^1S_0$, $^3S_1$ $C_0$'s, LO OPE  & \\
\hline
\ \ $Q^{-1/2}$ & $^3P_J$, $^3D_J$ $C_0$'s & \\
\ \ $Q^0$ & $^1S_0$ $C_2$ & \\
\ \ $Q^{1/2}$ & $^3S_1$  $C_2$ & \\
\ \ $Q^{5/4}$ & & $^1S_0$--$^3S_1$ $C_{D0}$ OPE\\
\ \ $Q^{3/2}$ & $^3P_J$, $^3D_J$ $C_2$'s & \\
\ \ $Q^{7/4}$ & & $^3P_0$--$^1S_0$ $C_{D0}$ OPE\\
\ \ $Q^2$ &\ \ $^1S_0$ $C_4$, $^1P_1$ $C_0$, NLO OPE, LO TPE \ \ & 
\ \ $^3P_1$--$^3S_1$ $C_{D0}$ OPE\\
\ \ $Q^{5/2}$ & $^3S_1$ $C_4$ &\ \  $^3P_J$, $^3D_J$ $C_{D0}$'s OPE\ \ \\
\ \ $Q^3$ & NLO TPE & LO 3N TPE \\
\ \ $Q^?$ & & $C_E$ \\
\hline
\end{tabular}
\end{center}
\caption{Orders of terms in the two- and three-nucleon effective
potentials for waves with $L\leq 2$. The leading coefficient in each 
interaction is labelled by the subscript 0, a subleading one (with one
power of the energy or two derivatives) by the subscript 2, and so on.}
\end{table}

The results of the various RG analyses of nuclear forces that I have outlined 
above are summarised in Table 1. This shows all two- and three-body forces 
up to order $Q^3$, which would correspond to N$^2$LO in Weinberg's power
counting. As already discussed, the ones of order $Q^{-1}$ must be iterated,
and it is probably most convenient to iterate also those of order $Q^{-1/2}$.
The remainder ought to be treated as perturbations.\footnote{In practice, 
treating parts of the potential in perturbation theory
does not mesh well with standard many-body methods, which treat the whole
potential to all orders. Nonetheless, we can still use these methods 
with potentials from EFTs, provided we are careful. The 
unrenormalised divergences that could destroy the power counting must be 
kept small. This can be achieved by using a cutoff $\Lambda$ that is well 
below the $\Lambda_0$, the breakdown scale of the EFT. The price for doing 
this is the introduction of large artefacts $\propto (Q/\Lambda)^{n}$, 
so that the radius of convergence of the EFT becomes $\Lambda$ rather than 
$\Lambda_0$. Keeping these artefacts small leaves us with only a narrow 
window of acceptable cutoffs, just below $\Lambda_0$.}

Among the two-body interactions, the leading contact interactions in the
$^3D_J$ waves, the subleading ones in the $P$ and $D$ waves, and the $S$-wave
$C_4$'s would be absent from the corresponding N$^2$LO potential in Weinberg 
counting. However all but the subleading $D$-wave interactions are present in 
the widely used N$^3$LO potential.

The three-body OPE forces are more interesting. Only the $S$-wave $C_{D0}$ 
term is included in the state-of-the-art (N$^2$LO) three-body 
interaction. The promotion of the pion couplings for other channels
could be of practical importance since $P$ waves have been implicated in 
several observables that cannot be described by the current three-body 
forces (see e.g.~\cite{ma09}).

\section*{Acknowledgments}

I thank the INT, Seattle for its hospitality in April--June 2009, and the 
organisers of the program INT-09-1 ``Effective field theories and the 
many-body problem'' for forcing me to organise my thoughts on these issues.

\end{document}